\documentclass[12pt]{article}
\newcommand{ \R} {\mbox{\rm I$\!$R}}
\newcommand{ \C} {\mbox{\rm I$\!$C}}
\begin{document}

\title{Locally Anisotropic Supergravity and\\
Gauge Gravity on Noncommutative Spaces}
\author{S. I.\ Vacaru $^{*}$, I. A.\ Chiosa \thanks{
e--mail: svacaru@phys.asm.md, vacaru@lises.asm.md, sergiu$_{-}$%
vacaru@yahoo.com }\ , and Nadejda A. Vicol $^\diamondsuit$ \quad \\
\\
{\small * Institute of Applied Physics, Academy of Sciences, }\\
{\small  Academy str. 5, Chi\c sin\v au MD2028, Republic of Moldova }\\
{\small $\diamondsuit$ Faculty of Mathematics and Informatics, gr. 33 MI,}\\
{\small State University of Moldova,  Mateevici str. 60,
 Chi\c sin\v au MD2009}}
\date{November 24, 2000}
\maketitle

\begin{abstract}
We outline the the geometry of locally anisotropic (la) superspaces and
la--supergravity. The approach is backgrounded on the method of anholonomic
superframes with associated nonlinear connection structure. Following the
formalism of enveloping algebras and star product calculus we propose a
model of gauge la--gravity on noncommutative spaces. The corresponding
Seiberg--Witten maps are established which allow the definition of dynamics
for a finite number of gravitational gauge field components on
noncommutative spaces.
\end{abstract}


\section{Introduction}

Locally anisotropic supergravity was developed as a model of supergravity
with anholonomic superframes and associated nonlinear connection
(N--connection) structure \cite{vs}. This model contain as particular cases
supersymmetric Kaluza--Klein and generalized Lagrange and/or Finsler
gravities and for nontrivial curvatures the N--connection describes
splittings from higher to lower dimensions of (super) spaces and generic
anholonomic local anisotropies.

In order to avoid the problem of formulation of gauge theories on
noncommutative spaces \cite{cds,sw,js,mssw} with Lie algebra valued
infinitesimal transformations and with Lie algebra valued gauge fields the
authors of \cite{jssw} suggested to use enveloping algebras of the Lie
algebras for setting this type of gauge theories and showed that in spite of
the fact that such enveloping algebras are infinite--dimensional one can
restrict them in a way that it would be a dependence on the Lie algebra
valued parameters and the Lie algebra valued gauge fields and their
spacetime derivatives only.

A still presented drawback of noncommutative geometry and physics is that
there is not yet formulated a generally accepted approach to interactions of
elementary particles coupled to gravity. There are improved Connes--Lott and
Chamsedine--Connes models of nocommutative geometry \cite{cl} which yielded
action functionals typing together the gravitational and Yang--Mills
interactions and gauge bosons the Higgs sector (see also the approaches \cite
{hawkins} and \cite{majid}).

In this paper we outline the geometry of locally anisotropoc supergravity
and follow the method of restricted enveloping algebras \cite{js,jssw} and
construct gauge gravitational theories by stating corresponding structures
with semisimple or nonsemisimple Lie algebras and their extensions. We
consider power series of generators for the affine and non linear realized
de Sitter gauge groups and compute the coefficient functions of all the
higher powers of the generators of the gauge group which are functions of
the coefficients of the first power. Such constructions are based on the
Seiberg--Witten map \cite{sw} and on the formalism of $*$--product
formulation of the algebra \cite{w} when for functional objects, being
functions of commuting variables, there are associated some algebraic
noncommutative properties encoded in the $*$--product. The concept of gauge
gravity theory on noncommutative spaces is introduced in a geometric manner
\cite{mssw} by defining the covariant coordinates without speaking about
derivatives and this formalism was developed for quantum planes \cite{wz}.
We prove the existence for noncommutative spaces of gauge models of gravity
which agrees with usual gauge gravity theories \cite{vn} being equivalent,
or extending, the general relativity theory (see works \cite{pd,ts} for
locally isotropic spaces and corresponding reformulations and
generalizations respectively for anholonomic frames \cite{vd} and locally
anisotropic (super) spaces \cite{vg}) in the limit of commuting spaces.

\section{Locally Anisotropic Supergravity}

Let us consider a vector superbundle (vs--bundle) $\widetilde{E}$ over a
supermanifold (s--manifold)$\widetilde{M}$ with surjective projection $\pi
_E:\widetilde{E}\rightarrow \widetilde{M}$ (for simplicity, all
constructions are locally trivial). The local supersymmetric coordinates
(s--coordinates) on $\widetilde{E}$ and $\widetilde{M}$ are denoted
respectively $u=(x,y)=\{u^\alpha =\left( x^I,y^A\right) ,$ where $%
x=\{x^I=\left( x^i,x^{\widehat{i}}\right) \}$ are (even,odd) coordinates on $%
\widetilde{M}$ and $y=\{y^A=\left( y^a,y^{\widehat{a}}\right) \}$ are
(even,odd) coordinates in fibers of $\pi _E$ (indices run values defined by
even and odd dimensions of corresponding submanifolds). Latin s--indices $%
I,J,K,L,M,...$ and $A,B,C,D,...$ will be used respectively for base and
fiber components.

A nonlinear connection (N--connection) structure which defines a global
decomposition of $T\widetilde{E}$ into horizontal, $H\widetilde{E},$ and
vertical parts, $V\widetilde{E},$%
\begin{equation}
\label{nc}
N:T\widetilde{E}=H\widetilde{E}\oplus V\widetilde{E}.
\end{equation}
The coefficients of a N--connection $N_I^A\left( u\right) $ determin the
locally adapted s--frame (basis, in brief la--frame)
\begin{equation}
\label{laf}
\delta _\alpha =\delta /\delta u^\alpha =\left( \delta _I=\delta /\delta
x^I=\partial _I-N_I^B\left( u\right) \partial _B,\partial _A\right) ,
\end{equation}
where $\partial _I=\partial /\partial x^I,\partial _A=\partial /\partial y^A$
are partial s--derivatives, and the dual s--frame
\begin{equation}
\label{dlaf}
\delta ^\alpha =\delta u^\alpha =\left( d^I=\delta x^I=dx^I,\delta ^A=\delta
y^A=dy^A+N_I^A\left( u\right) dx^I\right) .
\end{equation}

The s--frame 
(\ref{laf}) is anholonomic
$$
[\delta _J,\delta _K\}=\delta _J\delta _K-(-)^{|JK|}\delta _K\delta
_J=\Omega _{\ JK}^A\partial _A,
$$
where $|JK|=|J|\cdot |K|$ is defined by the parity of indices and we write $%
(-)^{|JK|}$ instead $(-1)^{|JK|},$ with anholonomy coefficients coinciding
with the N-connection curvature%
$$
\Omega _{\ JK}^A=\delta _KN_J^A-(-)^{|JK|}\delta _JN_K^A.
$$

The geometrical objects on $\widetilde{E}$ are given with respect to
la--basis 
(\ref{laf}) and 
(\ref{dlaf}) or their tensor products and called ds--tensors,
ds--connections (for some additional linear connections), d--spinors and so
on. For instance, a metric ds--tensor is written
\begin{equation}
\label{dm}
\widetilde{g}=g_{\alpha \beta }\delta ^\alpha \otimes \delta ^\beta
=g_{IJ}d^I\otimes d^J+g_{AB}\delta ^A\otimes \delta ^B.
\end{equation}

The Lagrange and Finsler ds--metrics can be modelled on a locally
an\-iso\-trop\-ic superspace  if vs--bundle $\widetilde{E}$ over a
s--manifold $\widetilde{M}$ is substituted by the tangent s--bundle $T%
\widetilde{M}$ and the coefficients of ds--metric 
(\ref{dm}) are taken respectively%
$$
g_{IJ}(u)=\frac 12\frac{\partial ^2{\cal L}(u)}{\partial y^I\partial y^L}%
\mbox{ and }g_{IJ}(u)=\frac 12\frac{\partial ^2{F}^2(u)}{\partial
y^I\partial y^L}
$$
where the s--Lagrangian ${\cal L:}T\widetilde{M}\rightarrow \Lambda $ is a
s--differentiable function on $T\widetilde{M},$ and ${F}$ is a Finsler
s--metric function on $T\widetilde{M}.$

A linear distinguished connection $D,$ d--connection, in sv--bundle $%
\widetilde{E}$ is a linear connection which preserves by parallelism the
horizontal (h) and vertical (v) distribution 
(\ref{nc}).

A d--connection $D\Gamma =\{\Gamma _{\ \beta \gamma }^\alpha =\left( L,%
\widetilde{L} ,\widetilde{C},C\right) \}, $ is determined by its invariant
hh-, hv-, vh- and vv--components, where
\begin{eqnarray}
D_{\left( \delta _K\right) }\delta _J &= &L_{~JK}^I\left( u\right) \delta
_I,\ D_{\left( \delta _K\right) }\partial _B=
L_{~BK}^A\left( u\right) \partial_A,  \label{dc} 
\\
D_{\left( \partial _C\right) }\delta _J &= &C_{~JC}^I\left( u\right) \delta
_I,\ D_{\left( \partial _C\right) }\partial _B=C_{~BC}^A\left( u\right)
\partial _A. \nonumber
\end{eqnarray}

There is a canonical d--connection $~^{(c)}\Gamma $ defined by the
coefficients of d--metric 
(\ref{dm}) and of N--connection and satisfying the metricity condition $D
\tilde g =0,$%
\begin{eqnarray}
^{(c)}L_{\ JK}^I &=&
\frac 12g^{IH}\left( \delta _Kg_{HJ}+\delta _Jg_{HK}-\delta
_Hg_{JK}\right) , \nonumber \\
^{(c)}L_{\ BK}^A &=&\partial _BN_K^A+\frac 12h^{AC}\left( \delta
_KH_{BC}-(\partial _BN_K^D)h_{DC}-(\partial _CN_K^D)h_{DB}\right) ,
\nonumber \\
^{(c)}C_{\ JC}^I &=&\frac 12g^{IK}\partial _Cg_{JK},~^{(c)}C_{\ BC}^A=\frac
12h^{AD}\left( \partial _Ch_{DB}+\partial _Bh_{DC}-\partial _Dh_{BC}\right).
\nonumber \end{eqnarray}

The torsion $T_{\ \beta \gamma }^\alpha $ of a d--connection, $T\left(
X,Y\right) =[X,DY\}-[X,Y\}, $ where $X$ and $Y$ are ds--vectors and by $%
[...\}$ we denote the s--anticommutator, is decomposed into hv--invariant
ds--torsions 
\begin{eqnarray}
hT\left( \delta _K,\delta _J\right)& = &
T_{\ JK}^I\delta _I,~vT\left( \delta
_K,\delta _J\right) =\widetilde{T}_{\ JK}^A\delta _I,~hT\left( \partial
_A,\delta _J\right) =\widetilde{P}_{\ JA}^I\delta _I, \nonumber \\
vT\left( \partial _B,\delta _J\right) & = &
P_{\ JB}^A\partial _A,\quad vT\left(\partial _C,\partial _B\right) =
S_{\ BC}^A\partial _A, \nonumber
\end{eqnarray}
with coefficients 
\begin{eqnarray}
T_{~JK}^I & = & L_{~JK}^I-(-)^{|JK|}L_{~KJ}^I,~\widetilde{T}_{\ JK}^A=\delta
_KN_J^A-(-)^{|KJ|}\delta _JN_K^A,~\label{dt} 
 \\
\widetilde{P}_{\ JA}^I & = & C_{\ JA}^I,\ P_{~JB}^A=\partial _BN_J^A-L_{\
BJ}^A,~S_{\ BC}^A=C_{\ BC}^A-(-)^{|BC|}C_{\ CB}^A. \nonumber
\end{eqnarray}
The even and odd components of ds--torsions 
(\ref{dt}) can be specified in explicit form by using decompositions of
indices into even and odd parts, $I=(i,\widehat{i)},A=\left( a,\widehat{a}%
\right) $ and so on.

The curvature $R_{\ \beta \gamma \tau }^\alpha $ of a d--connection, $%
R\left( X,Y\right) Z=D_{[X}D_{Y\}}Z-D_{[X,Y\}}Z, $ where $X,Y,Z$ are
ds--vectors, splits into hv--invariant ds--torsions%
\begin{eqnarray}
R\left( \delta _K,\delta _J\right) \delta _H &= & R_{~HJK}^I\delta _I,
~R\left(\delta _K,\delta _J\right) \partial _B=R_{~BJK}^A\partial _A,
\label{dcurvc} 
 \\
~R\left(\partial _C,\delta _K\right) \delta _J&=&
\widetilde{P}_{~JKC}^I\delta _I,\
R\left( \partial _C,\delta _K\right) \partial _B  =
 P_{~BKC}^A, \nonumber \\
R\left(\partial _C,\partial _B\right) \delta _J &=&
 \widetilde{S}_{~JBC}^I\delta_I,~R\left( \partial _D,
\partial _C\right) \partial _B=S_{~BCD}^A\partial _A
\nonumber
\end{eqnarray}
where the coefficients are computed%
\begin{eqnarray}
R_{\ MJK}^I&=&\delta _{[K}L_{\ |M|J\}}^I+
L_{\ MJ}^WL_{\ WK}^I-\left( -\right)^{|KJ|}L_{\ MK}^WL_{\ WJ}^I
+C_{\ KA}^IW_{\ JK}^A, \nonumber \\
\widehat{R}_{\ BJK}^A &=& \delta _{[K}L_{\ |B|J\}}^A+
L_{\ BJ}^CL_{\ CK}^A-\left(-\right) ^{|KJ|}L_{\ BK}^CL_{\ CJ}^A+
C_{\ BC}^AW_{\ JK}^C, \nonumber \\
\widetilde{S}_{\ JBC}^I & =& \partial _CC_{\ JB}^I-
\left( -\right)^{|BC|}\partial _BC_{\ JC}^I+C_{\ JB}^HC_{\ HC}^I-
\left( -\right)^{|BC|}C_{\ JC}^HC_{\ HB}^I, \nonumber \\
S_{\ BCD}^A & = &\partial _DC_{\ BC}^A-
\left( -\right) ^{|CD|}\partial _CC_{\ BD}^A+C_{\ BC}^EC_{\ ED}^A-
\left( -\right) ^{|CD|}C_{\ BD}^EC_{\ EC}^A, \nonumber \\
\widetilde{P}_{\ JKA}^I &= &\partial _AL_{\ JK}^I-C_{\ JA|K}^I+
C_{\ JB}^{I\,}P_{\ KA}^B, \nonumber \\
P_{\ BKC}^A &=& \partial _CL_{\ BK}^A-C_{\ BC|K}^A+C_{\ BD}^AP_{\ KC}^D,
\nonumber
\end{eqnarray}
where, for instance,%
\begin{eqnarray}
\delta _{[K}L_{\ |M|J\}}^I &=&\delta _KL_{\ MJ}^I-
\left( -\right) ^{|KJ|}\delta_JL_{\ MK}^I, \nonumber \\
C_{\ JA|K}^I &=& \delta _KC_{\ JA}^I+L_{\ MK}^IC_{\ IA}^M-
L_{\ JK}^MC_{\ MA}^I-L_{\ AK}^BC_{\ JB}^I. \nonumber
\end{eqnarray}
The even and odd components of ds--curvatures are computed by splitting
indices into even and odd parts.

The torsion and curvature of a d--connection $D$ on a sv--bundle satisfy the
identities%
\begin{eqnarray}
\sum\limits_{SC}\left[ (D_XT)(Y,Z)-R(X,Y)Z+T(T(X,Y),Z)\right] &= &0,
\nonumber \\
\sum\limits_{SC}\left[ (D_XR)(U,Y,Z)-R(T(X,Y),Z)U\right] &= &0,
\nonumber
\end{eqnarray}
where $\sum_{SC}$ means supersymmetric cyclic sums over ds--vectors $X,Y,Z$
and $U,$ from which the generalized Bianchi and Ricci identities follow
[1-3].

The Ricci ds--tensor $R_{\beta \gamma }=R_{\ \beta \gamma \alpha }^\alpha $
has hv--invariant components%
\begin{eqnarray}
R_{IJ} &= &R_{~IJK}^K,
~R_{IA}=-~^{(2)}P_{IA}=-(-)^{|KA|}\widetilde{P}_{\ IKA}^K,
\label{dsricci} 
\\
R_{AI} & =& ~^{(1)}P_{AI}=P_{\ AIB}^B,~R_{AB}=S_{~ABC}^C=S_{AB}. \nonumber
\end{eqnarray}

If a ds--metric 
(\ref{dm}) is defined on $\widetilde{E},$ we can introduce the
supersymmetric scalar curvature 
$$
\widehat{R}=g^{\alpha \beta }R_{\alpha \beta }=R+S, 
$$
where $R=g^{IJ}R_{IJ}$ and $S=h^{AB}S_{AB}.$

The simplest model of locally anisotropic supergravity (la--supergravity)
was constructed by postulating a variant of supersymmetric Einstein--Cartan
theory on LAS--space $\widetilde{E},$ which in invariant hv--components has
the fundamental s--field equations%
\begin{eqnarray}
R_{IJ}&-&\frac 12\left( R+S-\lambda \right) g_{IJ}=k_1 \Upsilon
_{IJ},~^{(1)}P_{AI}=k_1 \Upsilon _{AI},\label{ect} 
 \\
S_{AB}&-&\frac 12\left( R+S-\lambda \right) h_{AB}=k_1 \Upsilon
_{AB},~^{(2)}P_{IA}=-k_1 \Upsilon _{IA}, \nonumber
\end{eqnarray}
and 
$$
T_{\ \beta \gamma }^\alpha +\delta _\beta ^\alpha T_{\ \gamma \tau }^\tau
-(-)^{|\beta \gamma |}\delta _\gamma ^\alpha T_{\ \beta \tau }^\tau =k_2
Q_{\ \beta \gamma }^\alpha , 
$$
where $\lambda$ is the cosmological constant, $k_{1,2}$ are respective
interaction constants $\Upsilon _{\alpha \beta }$ is the energy--momentum
ds--tensor and $Q_{\ \beta \gamma }^\alpha $ is defined by the
supersymmetric spin--density.

The bulk of theories of locally isotropic s--gravity are formulated as gauge
supersymmetric models based on supervielbein formalism. Similar approaches
to la--supergravity on vs--bundles can be developed by considering arbitrary
s--frames $B_{\underline{\alpha }}\left( u\right) =\left( B_{\underline{I}%
}\left( u\right) ,B_{\underline{C}}\left( u\right) \right) $adapted to the
N--connection structure on a vs-bundle $\widetilde{E}=\widetilde{E}^{m,l}$
over s--manifold $\widetilde{M}=\widetilde{M}^{n,k}$ where $(m,l)$ and $%
(n,k) $ are respective (even, odd) dimensions of s--manifolds. A s--frame $%
B_{\underline{\alpha }}\left( u\right) $ is related with a standard
la--frame 
(\ref{laf}) via transforms $
\delta _\alpha =A_\alpha ^{~\underline{\alpha }}\left( u\right) B_{%
\underline{\alpha }}\left( u\right) ,$ 
where s--matrices $A_\alpha ^{~\underline{\alpha }}\left( u\right) =\left( 
\begin{array}{cc}
A_I^{~\underline{I}} & 0 \\ 
0 & A_C^{~\underline{C}} 
\end{array}
\right) $ take values into a super Lie group $GL_{n,k}^{m,l}\left( \Lambda
\right) =GL\left( n,k,\Lambda \right) \oplus GL\left( m,l,\Lambda \right) $
(on superspaces the graded Grassmann algebra with Euclidean topology,
denoted by $\Lambda ,$ substitutes the real and complex number fields).

We denote by $LN\left( \widetilde{E}\right) $ the set of all adapted to
N--connection s--frames in all points of vs--bundle $\widetilde{E}$ and
consider the s--bundle of linear adapted s--frames on $\widetilde{E}$
defined as the principal s--bundle%
$$
{\cal L}N\left( \widetilde{E}\right) =\left( LN\left( \widetilde{E}\right)
,\pi _L:LN\left( \widetilde{E}\right) \rightarrow \widetilde{E}%
,GL_{n,k}^{m,l}\left( \Lambda \right) \right) , 
$$
for a surjective s--map $\pi _L.$ The canonical basis of standard
distinguished s--generators $I_{\widehat{\alpha }}\rightarrow I_{\underline{%
\beta }}^{\underline{\alpha }}=\left( 
\begin{array}{cc}
I_{\underline{J}}^{\underline{I}} & 0 \\ 
0 & I_{\underline{B}}^{\underline{A}} 
\end{array}
\right) $ for the super Lie algebra ${\cal G}L_{n,k}^{m,l}\left( \Lambda
\right) $ of the structural s--group $GL_{n,k}^{m,l}\left( \Lambda \right) $
satisfy s--commutation rules $[I_{\widehat{\alpha }},I_{\widehat{\beta }%
}\}=f_{\widehat{\alpha }\widehat{\beta }}^{\quad \widehat{\gamma }}I_{%
\widehat{\gamma }}. $ On ${\cal L}N\left( \widetilde{E}\right) $ we consider
the d--connection 1--form%
$$
{\cal F}=\Gamma _{\ \underline{\beta }\gamma }^{\underline{\alpha }}\left(
u\right) I_{\underline{\alpha }}^{\underline{\beta }}\delta u^\gamma , 
$$
where 
\begin{equation}
\label{dclb}
\Gamma _{\ \underline{\beta }\gamma }^{\underline{\alpha }}\left( u\right)
=A_\alpha ^{\ \underline{\alpha }}A_{\ \underline{\beta }}^\beta \Gamma _{\
\beta \gamma }^\alpha +A_\beta ^{\ \underline{\alpha }}\delta _\gamma A_{\ 
\underline{\beta }}^\beta , 
\end{equation}
$\Gamma _{\ \beta \gamma }^\alpha $ are the components of canonical variant
of d--connection 
(\ref{dc}) and the s--matrix $A_{\ \underline{\beta }}^\beta $ is inverse to 
$A_\alpha ^{\ \underline{\alpha }}.$

The curvature ${\cal B}$ of the d--connection 
(\ref{dclb}) 
\begin{equation}
\label{dcurvlb}
{\cal B}=\delta {\cal F}+{\cal F}\wedge {\cal F}=R_{\ \underline{\alpha }%
\gamma \tau }^{\underline{\beta }}I_{\underline{\beta }}^{\underline{\alpha }%
}\delta u^\gamma \wedge \delta u^\tau 
\end{equation}
has the coefficients 
 $
R_{\ \underline{\alpha }\gamma \tau }^{\underline{\beta }}=A_{\ \underline{%
\alpha }}^\alpha \left( u\right) A_\beta ^{\ \underline{\beta }}\left(
u\right) R_{\ \alpha \gamma \tau }^\beta , $
where $R_{\ \alpha \gamma \tau }^\beta $ are defined by ds--curvatures 
(\ref{dcurvc}).

Aside from ${\cal L}N\left( \widetilde{E}\right) $ with vs--bundle $%
\widetilde{E}$ is naturally related another s--bundle, the bundle of adapted
to N--connection affine s--frames 
$$
{\cal A}N\left( \widetilde{E}\right) =\left( AN\left( \widetilde{E}\right)
,\pi _A:AN\left( \widetilde{E}\right) \rightarrow \widetilde{E}%
,AF_{n,k}^{m,l}\left( \Lambda \right) \right) , 
$$
with the affine strucural s--group $AF_{n,k}^{m,l}\left( \Lambda \right)
=GL_{n,k}^{m,l}\left( \Lambda \right) \odot \Lambda ^{n,k}\oplus \Lambda
^{m,l}.$

The d--connection ${\cal F}$ 
(\ref{dclb}) in ${\cal L}N\left( \widetilde{E}\right) $ induces in a linear
Cartan d--con\-nec\-ti\-on $\overline{{\cal F}}=\left( {\cal F},\chi \right)
,$ in ${\cal A}N\left( \widetilde{E}\right) ,$where $\chi =e_{\underline{%
\alpha }}\otimes A_\alpha ^{~\underline{\alpha }}\left( u\right) \delta
u^\alpha ,$ $e_{\underline{\alpha }}$ is the standard basis in $\Lambda
^{n,k}\oplus \Lambda ^{m,l},$ and, in consequence, the curvature ${\cal B}$ 
(\ref{dcurvlb}) in ${\cal L}N\left( \widetilde{E}\right) $ induces the pair
(curvature, torsion) $\overline{{\cal B}}=\left( {\cal B,T}\right) $ in $%
{\cal A}N\left( \widetilde{E}\right) ,$where 
$$
{\cal T}=\delta \chi +[{\cal F}\wedge \gamma \}={\cal T}_{\ \beta \gamma }^{%
\underline{\alpha }}e_{\underline{\alpha }}\delta u^\beta \wedge \delta
u^\gamma , 
$$
when ${\cal T}_{\ \beta \gamma }^{\underline{\alpha }}=A_\alpha ^{~%
\underline{\alpha }}T_{\ \beta \gamma }^\alpha \,$ is defined by the
coefficients of d--torsions 
(\ref{dt}) .

By using the ds--metric 
(\ref{dm}) in $\widetilde{E}$ one defines the (dual for s--forms) Hodge
operator $*_{\widetilde{g}}.$ Let the operator $*_{\widetilde{g}}^{-1}$ be
inverse to $*_{\widetilde{g}}$ and $\widehat{\delta }_{\widetilde{g}}$ be
the adjoint to the absolute derivation $\widehat{\delta }$ (associated to
the scalar product of ds--forms) specified for (r,s)--forms $\widehat{\delta 
}_{\widetilde{g}}=\left( -1\right) ^{r+s}*_{\widetilde{g}}^{-1}\circ 
\widehat{\delta }\circ *_{\widetilde{g}}.$

The supersymmetric variant of the Killing form of the s--group $%
AF_{n,k}^{m,l}\left( \Lambda \right) $ is degenerate. In order to generate a
metric structure $\widetilde{g}_A$ in the total spaces of the s--bundle $%
{\cal A}N\left( \widetilde{E}\right) $ we use and auxiliary nondegenerate
bilinear s--form which gives rise to the possibility to define the Hodge
operator $*_{\widetilde{g}_A}$ and $\widehat{\delta }_{\widetilde{g}A}.$
Applying the operator of horizontal projection $\widehat{H}$ one defines the
operator $\triangle \doteq \widehat{H}\circ \widehat{\delta }_{\widetilde{g}%
A}$ which does not depend on components of auxiliary biliniar s--form in the
fiber.

Following an abstract geometric calculus, by using operators $*_{\widetilde{g%
}},*_{\widetilde{g}_A},\widehat{\delta }_{\widetilde{g}},\widehat{\delta }_{%
\widetilde{g}A}$ and $\triangle $ one computers 
\begin{equation}
\label{abder}
\triangle \overline{{\cal B}}=\left( \triangle {\cal B},{\cal R}t+{\cal R}i
\right) , 
\end{equation}
where the one s--forms 
\begin{eqnarray}
{\cal R}t & = &\widehat{\delta }_{\widetilde{g}}{\cal T}+
*_{\widetilde{g}}^{-1}[{\cal F},*_{\widetilde{g}}{\cal T}\}, \nonumber \\
{\cal R}i &
= &*_{\widetilde{g}}^{-1}[\chi ,*_{\widetilde{g}}{\cal B}%
\}=(-1)^{n+k+l+m}R_{\alpha \beta }g^{\alpha \widehat{\beta }}
e_{\widehat{\beta }}\delta u^\beta  \nonumber
\end{eqnarray}
are constructed respectively by using the ds--torsions 
(\ref{dt}) and Ricci ds--tensors 
(\ref{dsricci}).

Let us introduce the locally anisotropic supersymmetric matter source $%
\overline{{\cal J}}$ constructed by using the same formulas from 
(\ref{abder}) when instead of $R_{\alpha \beta }$ is taken $k_1(\Upsilon
_{\alpha \beta }-\frac 12g_{\alpha \beta }\Upsilon )-\lambda \left(
g_{\alpha \beta }-\frac 12g_{\alpha \beta }\delta _\tau ^\tau \right) .$ By
straightforward calculations we can proof [3,4] that the Yang--Mills
equations 
\begin{equation}
\label{ymeq}
\triangle \overline{{\cal B}}=\overline{{\cal J}} 
\end{equation}
for d--connection $\overline{{\cal F}}=\left( {\cal F},\chi \right) $ in
s--bundle ${\cal A}N\left( \widetilde{E}\right) ,\,$ projected on the base
s--manifold, are equivalent to the Einstein equations 
(\ref{ect}) on $\widetilde{E}.$ We emphasize that the equations 
(\ref{ymeq}) were introduced in a ''pure'' geometric manner by using
operators $*,\widehat{\delta }$ and the horizontal projection $\widehat{H}$
but such gauge s--field equations are not variational because of
degeneration of the Killing s--form. To construct a variational gauge like
supersymmetric la--supergravitational model is possible, for instance, by
considering a minimal extension of the gauge s--group $AF_{n,k}^{m,l}\left(
\Lambda \right) $ to the de Sitter s--group $S_{n,k}^{m,l}\left( \Lambda
\right) =SO_{n,k}^{m,l}\left( \Lambda \right) ,$ acting on the s--space $%
\Lambda _{n,k}^{m,l}\oplus \Lambda $ and formulating a nonlinear version of
de Sitter gauge s--gravity.

There are analyzed models of supergravity with generic local anisot\-ro\-py 
\cite{vs} when instead of s--field equations and constraints 
(\ref{ect}) there are considered an anholonomic generalization of the
Wess--Zumino supergravity and some variants induced in low energy limit from
superstring theory.
 The N--connection s--field allows us to model
generic la--interactions with dynamics and constraints induced by nontrivial
(not only via toroidal compactifications) from higher dimensions and this
results in a geometrical unification of the so--called generalized
Finsler--Kaluza--Klein theories.

\section{*--Products and Enveloping Algebras \protect\newline in
Noncommutative Spaces}

For a noncommutative space the coordinates ${\hat u}^i,$ $(i=1,...,N)$
satisfy some noncommutative relations of type 
\begin{equation}
\label{ncr}[{\hat u}^i,{\hat u}^j]=\left\{ 
\begin{array}{rcl}
& i\theta ^{ij}, & \theta ^{ij}\in 
\C,\mbox{ canonical structure; } \\  & if_k^{ij}{\hat u}^k, & f_k^{ij}\in 
\C,\mbox{ Lie structure; } \\  & iC_{kl}^{ij}{\hat u}^k{\hat u}^l, & 
C_{kl}^{ij}\in \C ,\mbox{ quantum plane structure} 
\end{array}
\right. 
\end{equation}
where $\C$ denotes the complex number field.

The noncommutative space is modeled as the associative algebra of $\C;$\
this algebra is freely generated by the coordinates modulo ideal ${\cal R}$
generated by the relations (one accepts formal power series)\ ${\cal A}_u=%
\C[[{\hat u}^1,...,{\hat u}^N]]/{\cal R.}$ One restricts attention \cite
{jssw} to algebras having the (so--called, Poincare--Birkhoff--Witt)
property that any element of ${\cal A}_u$ is defined by its coefficient
function and vice versa,%
$$
\widehat{f}=\sum\limits_{L=0}^\infty f_{i_1,...,i_L}:{\hat u}^{i_1}\ldots {%
\hat u}^{i_L}:\quad \mbox{ when }\widehat{f}\sim \left\{ f_i\right\} , 
$$
where $:{\hat u}^{i_1}\ldots {\hat u}^{i_L}:$ denotes that the basis
elements satisfy some prescribed order (for instance, the normal order $%
i_1\leq i_2\leq \ldots \leq i_L,$ or, another example, are totally
symmetric). The algebraic properties are all encoded in the so--called
diamond $(\diamond )$ product which is defined by 
$$
\widehat{f}\widehat{g}=\widehat{h}~\sim ~\left\{ f_i\right\} \diamond
\left\{ g_i\right\} =\left\{ h_i\right\} . 
$$

In the mentioned approach to every function $f(u)=f(u^1,\ldots ,u^N)$ of
commuting variables $u^1,\ldots ,u^N$ one associates an element of algebra $%
\widehat{f}$ when the commuting variables are substituted by anticommuting
ones, 
$$
f(u)=\sum f_{i_1\ldots i_L}u^1\cdots u^N\rightarrow \widehat{f}%
=\sum\limits_{L=0}^\infty f_{i_1,...,i_L}:{\hat u}^{i_1}\ldots {\hat u}%
^{i_L}: 
$$
when the $\diamond $--product leads to a bilinear $*$--product of functions
(see details in \cite{mssw})%
$$
\left\{ f_i\right\} \diamond \left\{ g_i\right\} =\left\{ h_i\right\} \sim
\left( f*g\right) \left( u\right) =h\left( u\right) . 
$$

The $*$--product is defined respectively for the cases (\ref{ncr}) 
$$
f*g=\left\{ 
\begin{array}{rcl}
\exp [{\frac i2}{\frac \partial {\partial u^i}}{\theta }^{ij}\frac \partial
{\partial {u^{\prime }}^j}]f(u)g(u^{\prime }){|}_{u^{\prime }\to u}, & 
\mbox{  canonical structure;} &  \\ 
\exp [\frac i2u^kg_k(i\frac \partial {\partial u^{\prime }},i\frac \partial
{\partial u^{\prime \prime }})]f(u^{\prime })g(u^{\prime \prime }){|}%
_{u^{\prime \prime }\to u}^{u^{\prime }\to u}, & \mbox{  Lie structure;} &  
\\ 
q^{{\frac 12}(-u^{\prime }{\frac \partial {\partial u^{\prime }}}v{\frac
\partial {\partial v}}+u{\frac \partial {\partial u}}v^{\prime }{\frac
\partial {\partial v^{\prime }}})}f(u,v)g(u^{\prime },v^{\prime }){|}%
_{v^{\prime }\to v}^{u^{\prime }\to u}, & \mbox{  quantum plane}, &  
\end{array}
\right. 
$$
where there are considered values of type%
\begin{eqnarray}
e^{ik_n\widehat{u}^n}e^{ip_{nl}\widehat{u}^n} &=&e^{i\{k_n+p_n+\frac
12g_n\left( k,p\right) \}\widehat{u}^n,}  \nonumber \\
g_n\left( k,p\right) & =& -k_ip_jf_{\ n}^{ij}+\frac 16k_ip_j\left(
p_k-k_k\right) f_{\ m}^{ij}f_{\ n}^{mk}+..., \label{gdecomp} \\
e^Ae^B &=&
e^{A+B+\frac 12[A,B]+\frac 1{12}\left( [A,[A,B]]+[B,[B,A]]\right)
}+... \nonumber
\end{eqnarray}
and for the coordinates on quantum (Manin) planes one holds the relation $%
uv=qvu.$

A non--abelian gauge theory on a noncommutative space is given by two
algebraic structures, the algebra ${\cal A}_u$ and a non--abelian Lie
algebra ${\cal A}_I$ of the gauge group with generators $I^1,...,I^S$ and
the relations 
\begin{equation}
\label{commutators1}[I^{\underline{s}},I^{\underline{p}}]=if_{~\underline{t}%
}^{\underline{s}\underline{p}}I^{\underline{t}}. 
\end{equation}
In this case both algebras are treated on the same footing and one denotes
the generating elements of the big algebra by $\widehat{u}^i,$%
\begin{eqnarray}
\widehat{z}^{\underline i} &=&
\{\widehat{u}^1,...,\widehat{u}^N,I^1,...,I^S\}, \nonumber \\
{\cal A}_z &=&\C[[\widehat{u}^1,...,\widehat{u}^{N+S}]]/{\cal R,}
\nonumber
\end{eqnarray}
and the $*$--product formalism is to be applied for the whole algebra ${\cal %
A}_z$ when there are considered functions of the commuting variables $u^i\
(i,j,k,...=1,...,N)$ and $I^{\underline{s}}\ (s,p,...=1,...,S).$

For instance, in the case of a canonical structure for the space variables $%
u^i$ we have 
\begin{equation}
\label{csp1}(F*G)(u)=e^{\frac i2\left( \theta ^{ij}\frac \partial {\partial
u^{\prime i}}\frac \partial {\partial u^{\prime \prime j}}+t^sg_s\left(
i\frac \partial {\partial t^{\prime }},i\frac \partial {\partial t^{\prime
\prime }}\right) \right) }F\left( u^{\prime },t^{\prime }\right) G\left(
u^{\prime \prime },t^{\prime \prime }\right) \mid _{t^{\prime }\rightarrow
t,t^{\prime \prime }\rightarrow t}^{u^{\prime }\rightarrow u,u^{\prime
\prime }\rightarrow u}. 
\end{equation}
This formalism was developed in \cite{jssw} for general Lie algebras. In
this paper we shall consider those cases when in the commuting limit one
obtains the gauge gravity and general relativity theories.

\section{Enveloping Algebras for \protect\newline Gravitational Gauge
Connections}

To define gauge gravity theories on noncommutative space we first introduce
gauge fields as elements the algebra ${\cal A}_u$ that form representation
of the generator $I$--algebra for the de Sitter gauge group. For commutative
spaces it is known \cite{pd,ts,vg} that an equivalent reexpression of the
Einstein theory as a gauge like theory implies, for both locally isotropic
and anisotropic spacetimes, the nonsemisimplicity of the gauge group, which
leads to a nonvariational theory in the total space of the bundle of locally
adapted affine frames (to this class one belong the gauge Poincare
theories;\ on metric--affine and gauge gravity models see original results
and reviews in \cite{ut}). By using auxililiary biliniear forms, instead of
degenerated Killing form for the affine structural group, on fiber spaces,
the gauge models of gravity can be formulated to be variational. After
projection on the base spacetime, for the so--called Cartan connection form,
the Yang--Mills equations transforms equivalently into the Einstein
equations for general relativity \cite{pd}. A variational gauge
gravitational theory can be also formulated by using a minimal extension of
the affine structural group ${{\cal A}f}_{3+1}\left( {\R}\right) $ to the de
Sitter gauge group $S_{10}=SO\left( 4+1\right) $ acting on ${\R}^{4+1}$
space. For cimplicity, in this paper we restrict our consideration only with
the even components  of frames, connections and curvatures of
  gauge la--supergavity outlined in previous section.

Let now consider a noncommutative space. In this case the gauge fields are
elements of the algebra $\widehat{\psi }\in {\cal A}_I^{(dS)}$ that form the
nonlinear representation of the de Sitter Lie algebra ${{\it s}o}_{\left(
\eta \right) }\left( 5\right) $ when the whole algebra is denoted ${\cal A}%
_z^{(dS)}.$ Under a nonlinear de Sitter transformation the elements
transform as follows%
$$
\delta \widehat{\psi }=i\widehat{\gamma }\widehat{\psi },\widehat{\psi }\in 
{\cal A}_u,\widehat{\gamma }\in {\cal A}_z^{(dS)}. 
$$
So, the action of the generators  on $\widehat{\psi }$ is
defined as this element is supposed to form a nonlinear representation of $%
{\cal A}_I^{(dS)}$ and, in consequence, $\delta \widehat{\psi }\in {\cal A}_u
$ despite $\widehat{\gamma }\in {\cal A}_z^{(dS)}.$ It should be emphasized
that independent of a representation the object $\widehat{\gamma }$ takes
values in enveloping de Sitter algebra and not in a Lie algebra as would be
for commuting spaces. The same holds for the connections that we introduce
(similarly to \cite{mssw}) in order to define covariant coordinates%
$$
\widehat{U}^\nu =\widehat{u}^v+\widehat{\Gamma }^\nu ,\widehat{\Gamma }^\nu
\in {\cal A}_z^{(dS)}. 
$$

The values $\widehat{U}^\nu \widehat{\psi }$ transforms covariantly, $\delta 
\widehat{U}^\nu \widehat{\psi }=i\widehat{\gamma }\widehat{U}^\nu \widehat{%
\psi },$ if and only if the connection $\widehat{\Gamma }^\nu $ satisfies
the transformation law of the enveloping nonlinear realized de Sitter
algerba,%
$$
\delta \widehat{\Gamma }^\nu \widehat{\psi }=-i[\widehat{u}^v,\widehat{%
\gamma }]+i[\widehat{\gamma },\widehat{\Gamma }^\nu ], 
$$
where $\delta \widehat{\Gamma }^\nu \in {\cal A}_z^{(dS)}.$ The enveloping
algebra--valued connection has infinitely many component fields.
Nevertheless, it was shown that all the component fields can be induced from
a Lie algebra--valued connection by a Seiberg--Witten map (\cite{sw,js,jssw}
and \cite{bsst} for $SO(n)$ and $Sp(n)).$ In this subsection we show that
similar constructions could be proposed for nonlinear realizations of de
Sitter algebra when the transformation of the connection is considered%
$$
\delta \widehat{\Gamma }^\nu =-i[u^\nu ,^{*}~\widehat{\gamma }]+i[\widehat{%
\gamma },^{*}~\widehat{\Gamma }^\nu ]. 
$$
For simplicity, we treat in more detail the canonical case with the star
product (\ref{csp1}). The first term in the variation $\delta \widehat{%
\Gamma }^\nu $ gives 
$$
-i[u^\nu ,^{*}~\widehat{\gamma }]=\theta ^{\nu \mu }\frac \partial {\partial
u^\mu }\gamma . 
$$
Assuming that the variation of $\widehat{\Gamma }^\nu =\theta ^{\nu \mu
}Q_\mu $ starts with a linear term in $\theta $ we have%
$$
\delta \widehat{\Gamma }^\nu =\theta ^{\nu \mu }\delta Q_\mu ,~\delta Q_\mu
=\frac \partial {\partial u^\mu }\gamma +i[\widehat{\gamma },^{*}~Q_\mu ]. 
$$
We follow the method of calculation from the papers \cite{mssw,jssw} and
expand the star product (\ref{csp1}) in $\theta $ but not in $g_a$ and find
to first order in $\theta ,$%
\begin{equation}\label{series}
\gamma = \gamma _{\underline{a}}^1I^{\underline{a}}+
\gamma _{\underline{a} \underline{b}}^1I^{\underline{a}}I^{\underline{b}}+..., 
 \mbox{ and } Q_\mu = q_{\mu ,\underline{a}}^1I^{\underline{a}}+
q_{\mu ,\underline{a} \underline{b}}^2I^{\underline{a}}I^{\underline{b}}+...
\end{equation}
where $\gamma _{\underline{a}}^1$ and $q_{\mu ,\underline{a}}^1$ are of
order zero in $\theta $ and $\gamma _{\underline{a}\underline{b}}^1$ and $%
q_{\mu ,\underline{a}\underline{b}}^2$ are of second order in $\theta .$ The
expansion in $I^{\underline{b}}$ leads to an expansion in $g_a$ of the $*$%
--product because the higher order $I^{\underline{b}}$--derivatives vanish.
For de Sitter case as $I^{\underline{b}}$ we take the generators,
 see commutators (\ref{commutators1}), with the corresponding de Sitter
structure constants $f_{~\underline{d}}^{\underline{b}\underline{c}}\simeq
f_{~\underline{\beta }}^{\underline{\alpha }\underline{\beta }}$ (in our
further identifications with spacetime objects like frames and connections
we shall use Greeck indices).

The result of calculation of variations of (\ref{series}), by using $g_a$ to
the order given in (\ref{gdecomp}), is%
\begin{eqnarray}
\delta q_{\mu ,\underline{a}}^1
 &=&\frac{\partial \gamma _{\underline{a}}^1}{\partial u^\mu }- %
f_{~\underline{a}}^{\underline{b}\underline{c}} %
\gamma _{\underline{b}}^1q_{\mu ,\underline{c}}^1, \nonumber \\  \delta Q_\tau &=& \theta ^{\mu \nu } %
\partial _\mu \gamma _{\underline{a}}^1\partial _\nu q_{\tau , \underline{b}}^1I^{\underline{a}}I^{\underline{b}}+...,  \nonumber \\ \delta q_{\mu ,\underline{a}\underline{b}}^2 &=&  %
\partial _\mu \gamma _{\underline{a}\underline{b}}^2 %
 -\theta ^{\nu \tau }\partial _\nu \gamma _{\underline{a}}^1 %
\partial _\tau q_{\mu ,\underline{b}}^1- %
2f_{~\underline{a}}^{\underline{b}\underline{c}} %
\{\gamma _{\underline{b}}^1q_{\mu ,\underline{c}\underline{d}}^2+ %
\gamma _{\underline{b}\underline{d}}^2q_{\mu ,\underline{c}}^1\}. %
 \nonumber
\end{eqnarray}

Next we introduce the objects $\varepsilon ,$ taking the values in de Sitter
Lie algebra and $W_\mu ,$ being enveloping de Sitter algebra valued,%
$$
\varepsilon =\gamma _{\underline{a}}^1I^{\underline{a}}\mbox{ and }W_\mu
=q_{\mu ,\underline{a}\underline{b}}^2I^{\underline{a}}I^{\underline{b}} 
$$
with the variation $\delta W_\mu $ satisfying the equation \cite{mssw,jssw} 
\begin{equation}
\label{vareq}\delta W_\mu =\partial _\mu (\gamma _{\underline{a}\underline{b}%
}^2I^{\underline{a}}I^{\underline{b}})-\frac 12\theta ^{\tau \lambda
}\{\partial _\tau \varepsilon ,\partial _\lambda q_\mu \}+i[\varepsilon
,W_\mu ]+i[(\gamma _{\underline{a}\underline{b}}^2I^{\underline{a}}I^{%
\underline{b}}),q_\nu ]. 
\end{equation}
The equation (\ref{vareq}) has the solution (found in \cite{mssw,sw})%
$$
\gamma _{\underline{a}\underline{b}}^2 = 
\frac 12\theta ^{\nu \mu }(\partial
_\nu \gamma _{\underline{a}}^1)q_{\mu ,\underline{b}}^1, 
 \mbox{ and }
q_{\mu ,\underline{a}\underline{b}}^2 = 
-\frac 12\theta ^{\nu \tau }q_{\nu ,\underline{a}}^1
\left( \partial _\tau q_{\mu ,\underline{b}}^1+ %
R_{\tau \mu ,\underline{b}}^1\right) 
$$
where $
R_{\tau \mu ,\underline{b}}^1=\partial _\tau q_{\mu ,\underline{b}%
}^1-\partial _\mu q_{\tau ,\underline{b}}^1+f_{~\underline{d}}^{\underline{e}%
\underline{c}}q_{\tau ,\underline{e}}^1q_{\mu ,\underline{e}}^1 $ 
can be identified with the coefficients ${\cal R}_{\quad \underline{\beta }%
\mu \nu }^{\underline{\alpha }}$ of de Sitter nonlinear gauge gravity
curvature if in the commutative limit $q_{\mu ,\underline{b}}^1\simeq \left( 
\begin{array}{cc}
\Gamma _{\quad \underline{\beta }}^{\underline{\alpha }} & l_0^{-1}\chi ^{
\underline{\alpha }} \\ l_0^{-1}\chi _{\underline{\beta }} & 0 
\end{array}
\right).$ 

The  presented procedure can be generalized to all higher powers of 
$\theta $ \cite{jssw}.

\section{Noncommutative Gauge Gravity Covariant Dynamics}

The constructions from the previous section are summarized by the conclusion
that the de Sitter algebra valued object $\varepsilon =\gamma _{\underline{a}%
}^1\left( u\right) I^{\underline{a}}$ determines all the terms in the
enveloping algebra%
$$
\gamma =\gamma _{\underline{a}}^1I^{\underline{a}}+\frac 14\theta ^{\nu \mu
}\partial _\nu \gamma _{\underline{a}}^1\ q_{\mu ,\underline{b}}^1\left( I^{%
\underline{a}}I^{\underline{b}}+I^{\underline{b}}I^{\underline{a}}\right)
+... 
$$
and the gauge transformations are defined by $\gamma _{\underline{a}%
}^1\left( u\right) $ and $q_{\mu ,\underline{b}}^1(u),$ when 
$$
\delta _{\gamma ^1}\psi =i\gamma \left( \gamma ^1,q_\mu ^1\right) *\psi . 
$$
For de Sitter enveloping algebras one holds the general formula for
compositions of two transformations%
$$
\delta _\gamma \delta _\varsigma -\delta _\varsigma \delta _\gamma =\delta
_{i(\varsigma *\gamma -\gamma *\varsigma )} 
$$
which holds also for the restricted transformations defined by $\gamma ^1,$%
$$
\delta _{\gamma ^1}\delta _{\varsigma ^1}-\delta _{\varsigma ^1}\delta
_{\gamma ^1}=\delta _{i(\varsigma ^1*\gamma ^1-\gamma ^1*\varsigma ^1)}. 
$$

Applying the formula (\ref{csp1}) we computer%
\begin{eqnarray}
[\gamma ,^{*}\zeta ]
 &=&i\gamma _{\underline{a}}^1\zeta _{\underline{b}}^1 %
f_{~\underline{c}}^{\underline{a}\underline{b}}I^{\underline{c}}+%
\frac i2\theta ^{\nu \mu }\{\partial _v\left( \gamma _{\underline{a}}^1%
\zeta _{\underline{b}}^1 %
f_{~\underline{c}}^{\underline{a}\underline{b}}\right) %
 q_{\mu ,\underline{c}} \nonumber \\  &{}&
+\left( \gamma _{\underline{a}}^1\partial _v\zeta _{\underline{b}}^1- %
\zeta _{\underline{a}}^1\partial _v\gamma _{\underline{b}}^1\right) %
 q_{\mu ,\underline{b}}f_{~\underline{c}}^{\underline{a}\underline{b}}+%
 2\partial _v %
\gamma _{\underline{a}}^1\partial _\mu \zeta _{\underline{b}}^1\} %
I^{\underline{d}}I^{\underline{c}}. \nonumber %
\end{eqnarray}
Such commutators could be used for definition of tensors \cite{mssw} 
\begin{equation}
\label{tensor1}\widehat{S}^{\mu \nu }=[\widehat{U}^\mu ,\widehat{U}^\nu ]-i%
\widehat{\theta }^{\mu \nu }, 
\end{equation}
where $\widehat{\theta }^{\mu \nu }$ is respectively stated for the
canonical, Lie and quantum plane structures. Under the general enveloping
algebra one holds the transform%
$$
\delta \widehat{S}^{\mu \nu }=i[\widehat{\gamma },\widehat{S}^{\mu \nu }]. 
$$
For instance, the canonical case is characterized by%
\begin{eqnarray}
S^{\mu \nu } &=&
i\theta ^{\mu \tau }\partial _\tau \Gamma ^\nu -i\theta ^{\nu
\tau }\partial _\tau \Gamma ^\mu +\Gamma ^\mu *\Gamma ^\nu -\Gamma ^\nu
*\Gamma ^\mu  \nonumber \\
& = &
\theta ^{\mu \tau }\theta ^{\nu \lambda }\{\partial _\tau Q_\lambda
-\partial _\lambda Q_\tau +Q_\tau *Q_\lambda -Q_\lambda *Q_\tau \}.
\nonumber
\end{eqnarray}
By introducing the gravitational gauge strength (curvature) 
\begin{equation}
\label{qcurv}R_{\tau \lambda }=\partial _\tau Q_\lambda -\partial _\lambda
Q_\tau +Q_\tau *Q_\lambda -Q_\lambda *Q_\tau , 
\end{equation}
which could be treated as a noncommutative extension of de Sitter nonlinear
gauge gravitational curvature one computers 
$$
R_{\tau \lambda ,\underline{a}}=R_{\tau \lambda ,\underline{a}}^1+\theta
^{\mu \nu }\{R_{\tau \mu ,\underline{a}}^1R_{\lambda \nu ,\underline{b}%
}^1-\frac 12q_{\mu ,\underline{a}}^1\left[ (D_\nu R_{\tau \lambda ,%
\underline{b}}^1)+\partial _\nu R_{\tau \lambda ,\underline{b}}^1\right]
\}I^{\underline{b}}, 
$$
where the gauge gravitation covariant derivative is introduced,%
$$
(D_\nu R_{\tau \lambda ,\underline{b}}^1)=\partial _\nu R_{\tau \lambda ,%
\underline{b}}^1+q_{\nu ,\underline{c}}R_{\tau \lambda ,\underline{d}}^1f_{~%
\underline{b}}^{\underline{c}\underline{d}}. 
$$
Following the gauge transformation laws for $\gamma $ and $q^1$ we find 
$$
\delta _{\gamma ^1}R_{\tau \lambda }^1=i\left[ \gamma ,^{*}R_{\tau \lambda
}^1\right] 
$$
with the restricted form of $\gamma .$

Such formulas were proved in references \cite{jssw,sw} for usual gauge
(nongravitational) fields. Here we reconsidered them for gravitational gauge
fields.

Following the nonlinear realization of de Sitter algebra and the $*$%
--formalism we can formulate a dynamics of noncommutative spaces.
Derivatives can be introduced in such a way that one does not obtain new
relations for the coordinates. In this case a Leibniz rule can be defined 
\cite{jssw} that 
$$
\widehat{\partial }_\mu \widehat{u}^\nu =\delta _\mu ^\nu +d_{\mu \sigma
}^{\nu \tau }\ \widehat{u}^\sigma \ \widehat{\partial }_\tau  
$$
where the coefficients $d_{\mu \sigma }^{\nu \tau }=\delta _\sigma ^\nu
\delta _\mu ^\tau $ are chosen to have not new relations when $\widehat{%
\partial }_\mu $ acts again to the right hand side. In consequence one holds
the $*$--derivative formulas 
$$
\partial _\tau *f=\frac \partial {\partial u^\tau }f+f*\partial _\tau , 
$$
$$
[\partial _l,^{*}(f*g)]=\left( [\partial _l,^{*}f]\right) *g+f*\left(
[\partial _l,^{*}g]\right)  
$$
and the Stokes theorem 
$
\int [\partial _l,f]=\int d^Nu[\partial _l,^{*}f]=\int d^Nu\frac \partial
{\partial u^l}f=0, $ 
where, for the canonical structure, the integral is defined,%
$$
\int \widehat{f}=\int d^Nuf\left( u^1,...,u^N\right) . 
$$

An action can be introduced by using such integrals. For instance, for a
tensor of type (\ref{tensor1}), when%
 $\delta \widehat{L}=i\left[ \widehat{\gamma },\widehat{L}\right] , $ 
we can define a gauge invariant action%
$$
W=\int d^Nu\ Tr\widehat{L},~\delta W=0, 
$$
were the trace has to be taken for the group generators.

For the nonlinear de Sitter gauge gravity a proper action is 
$$
L=\frac 14R_{\tau \lambda }R^{\tau \lambda }, 
$$
where $R_{\tau \lambda }$ is defined by the even part of (\ref{dcurvlb}).  In
this case the dynamic of noncommutative space is entirely formulated in the
framework of quantum field theory of gauge fields. The method works for
matter fields as well to restrictions to the general relativity theory (see
references \cite{ts,pd}).

\vskip0.1cm

{\bf Acknowledgement:} The first author (S. V.) is grateful to Organizes 
 of NATO ARW in Kiev, where the results of this work were communicated,
 for kind hospitality and support.


\begin{thebibliography}{99}
\bibitem{bsst}  L. Bonora, M. Schnabl, M. M. Sheikh--Jabbari and A.
Tomasiello, {\it Noncommutative SO(n) and Sp(n) gauge theories,}
hep--th/0006091.

\bibitem{cl}  A. Connes, J. Math. Phys. 36 (1995) 6194;\ 
A. H. Chamseddine, A. Connes, Phys. Rev. Letters 77 (1996) 4868;\ 
L. Carminati, B. Iochum, T. Sch\"ocer, {\it Noncommutative Yang--Mills and
noncommutative relativity: a bridge over troubled water,} hep--th/9706105.

\bibitem{cds}  A. Connes, M. R. Douglas and A. Schwarz, JHEP {\bf 9802}
(1998)\ 003,\ hep--th/0001203.

\bibitem{hawkins}  E. Hawkins, Comm. Math. Phys. 187 (1997) 471; gr--qc /
9605068;\ 
G. Landi, N. A. Viet, K. C. Wali, Phys. Lett. B326 (1994) 45; hep--th /
9402046;\ 
A, Sitarz, Class. Quant. Grav. 11 (1994) 2127; hep--th / 9401145;\ 
J. Madore and J. Mourad, Int. J. Mod. Phys. D3 (1994) 221.

\bibitem{js}  B. Jur\v co and P. Schupp, Eur. Phys. J. {\bf C 14}, (2000)
367, hep--th/0001032.

\bibitem{jssw}  B. Jur\v co, S. Schraml, P. Shupp and J. Wess, {\it %
Enveloping algebra valued gauge transformations for non--abelian gauge
groups on non--commutative spaces}, hep--th/0006246

\bibitem{mssw}  J. Madore, S. Schraml, P. Schupp and J. Wess, {\it Gauge
theory on noncommutative spaces,} Eur. Phys. J. {\bf C}, in press,
hep--th/0001203.

\bibitem{majid}  S. Majid, {\it Conceptual Issues for Noncommutative gravity
and algebras and finite sets,} math.QA / 0006152;\ 
A. H. Chamseddine, {\it Complexified gravity and noncommutative spaces,}
hep--th / 0005222; 
R. Kerner, {\it Noncommutative extensions of classical theories in physics, }
hep--th / 0004033.

\bibitem{pd}  D. A. Popov, Theor. Math. Phys. {\bf 24} (1975) 347;\ 
D. A. Popov and L. I. Dikhin, Doklady Akademii Nauk SSSR {\bf 245} (1975)
347 [in Russian].

\bibitem{sw}  N. Seiberg and E. Witten, JHEP {\bf 9909} (1999)\ 032,\
hep--th/9908142.

\bibitem{ts}  A. A. Tseytlin, Phys. Rev. D {\bf 26} (1982) 3327.

\bibitem{ut}  R. Utiyama, Phys. Rev. {\bf 101} (1956) 1597;\ 
V. N. Ponomariov, A. Barvinsky and Yu. N. Obukhov, {\it Geometrodynamical
Methods and the Gauge Approach to the Gravitational Interactions}
(Energoatomizdat, Moscow, 1985);\ 
E. W. Mielke, {\it Geometrodynamics of Gauge Fields -- on the Geometry of
Yang--Mills and Gravitational Gauge Theories} (Academic--Verlag, Berlin,
1987);\ 
F. Hehl, J. D. McGrea, E. W. Mielke and Y. Ne'eman, Phys. Rep. {\bf 258}
(1995) 1;\ 
H. Dehnen and E. Hitzer, {\it Int. J. Theor. Phys.}\ {\bf 34} (1995) 1981.

\bibitem{vs}  S. Vacaru, Nucl. Phys. B {\bf 434 }(1997) 590;\
gr--qc/9604016;\ hep--th / 9604194--9604196;\ Interactions, Strings and
Isotopies in Higher Order Anisotropic Superspaces (Hadronic Press, 1998).\ 

\bibitem{vn}  \ S.\ Vacaru, {\it Gauge and Eistein gravity from non--Abelian
gauge models on noncommutative spaces, }hep-th/0009163

\bibitem{vd}  S. Vacaru and H. Dehnen, {\it Locally Anisotropic Structures
and Nonlinear Connections in Einstein and Gauge Gravity,}\ gr--qc / 0009039.

\bibitem{vg}  S. Vacaru and Yu. Goncharenko, Int. J. Theor. Phys. {\bf 34}
(1995) 1955.

\bibitem{wz}  J. Wess and B. Zumino, Nucl. Phys. Proc. Suppl. {\bf 18B}
(1991) 302;\ 
J. Wess, in {\it Proceeding of the 38 International Universit\"atswochen
f\"ur Kern-- und Teilchenphysik}, no. 543 in Lect. Notes in Phys.,
Springer--Verlag, 2000, Schladming, January 1999, eds. H. Gusterer, H.
Grosse and L. Pitner;\ math--ph / 9910013.

\bibitem{w}  H. Weyl, Z. Physik {\bf 46} (1927) 1; {\it The theory of groups
and quantum mechanics} (Dover, New--York, 1931), translated from {\it %
Gruppentheorie and Quantenmechanik} (Hirzel Verlag, Leipzig, 1928);\ 
E. P. Wigner, Phys. Rev. {\bf 40} (1932) 749;\ 
J. E. Moyal, Proc. Cambridge Phil. Soc. {\bf 45} (1949) 99;\ 
F. Bayen, M. Flato, C. Fronsdal, A. Lichnerowicz, D. Sternheimer, Ann.
Physics\ {\bf 111} (1978) 61;\ 
M. Kontsevitch, {\it Deformation quantization of Poisson manifolds, I.}\
q--alg/9709040;\ 
D. Sternheimer, {\it Deformation Quantization: Twenty Years After,}\ math /
9809056.
\end{thebibliography}
\end{document}